\documentclass[12pt]{article}

\usepackage{times}
\usepackage{graphicx}
\newenvironment{sciabstract}{%
\begin{quote} \bf}
{\end{quote}}

\newcounter{lastnote}

\title{Inducing phase-locking and chaos in cellular oscillators by modulating the driving stimuli}

\author
{Mogens H. Jensen$^{1,\ast}$ and Sandeep Krishna$^{2}$\\
\\
\normalsize{$^{1}$Niels Bohr Institute, University of Copenhagen, 
Blegdamsvej 17, DK-2100 Copenhagen, Denmark,}\\
\normalsize{$^{2}$National Centre for Biological Sciences, GKVK Campus, Bellary Road, Bangalore, India.}
\\
\normalsize{$^\ast$To whom correspondence should be addressed; E-mail:  
mhjensen@nbi.dk}
}

\date{}

\begin{document}
\baselineskip24pt
\maketitle
\begin{sciabstract}
Inflammatory responses in eucaryotic cells are often associated
with oscillations in the nuclear-cytoplasmic translocation of the 
transcription factor NF-kB. In most laboratory realizations, the
oscillations are triggered by a cytokine stimulus,
like the tumor necrosis factor alpha, applied as 
a step change to a steady level.
Here we use a mathematical model to show that
an oscillatory external stimulus can synchronize the NF-kB oscillations
into states where the ratios of the internal to external frequency
are close to rational numbers. We predict a specific response diagram of the
TNF-driven NF-kB system which exhibits bands of synchronization known as ``Arnold tongues". 
Our model also suggests that when the amplitude of the external
stimulus exceeds a certain threshold there is the possibility of
coexistence of multiple different synchronized states and eventually
chaotic dynamics of the nuclear NF-kB concentration.
This could be used as a way of externally controlling immune response, DNA repair and 
apoptotic pathways.

\end{sciabstract}

The synchronization between two oscillating signals exhibits a 
surprisingly deep level of complexity \cite{Kurths}.
Already in 1876, the dutch physicist Huygens observed that two clocks
hanging on the wall tend to move in parallel after some time, i.e.,
they become synchronized \cite{CH}. 
Since then, such phenomena have been observed
in a variety of systems ranging from fluids to quantum mechanical 
devices \cite{JBB,Stavans,Martin,Gwinn,He,Gruner}. 
In recent years it has become increasingly clear that living 
organisms offer a bewildering fauna of oscillators, e.g.
cell cycles \cite{Ferrell}, circadian rhythms \cite{Lefranc}, 
embryo segmentation clocks \cite{Lykke}, 
calcium oscillations \cite{Goldbeter}, pace maker cells \cite{pacemaker}, 
protein responses \cite{Hoffmann02,Nelson04}, hormone secretion \cite{hormones},
and so on. A natural question therefore
is: do oscillators in cells, organs and tissues tend to
synchronize to each other or to external driving oscillations? 

\begin{figure}[h!]
\includegraphics[width=0.8\textwidth]{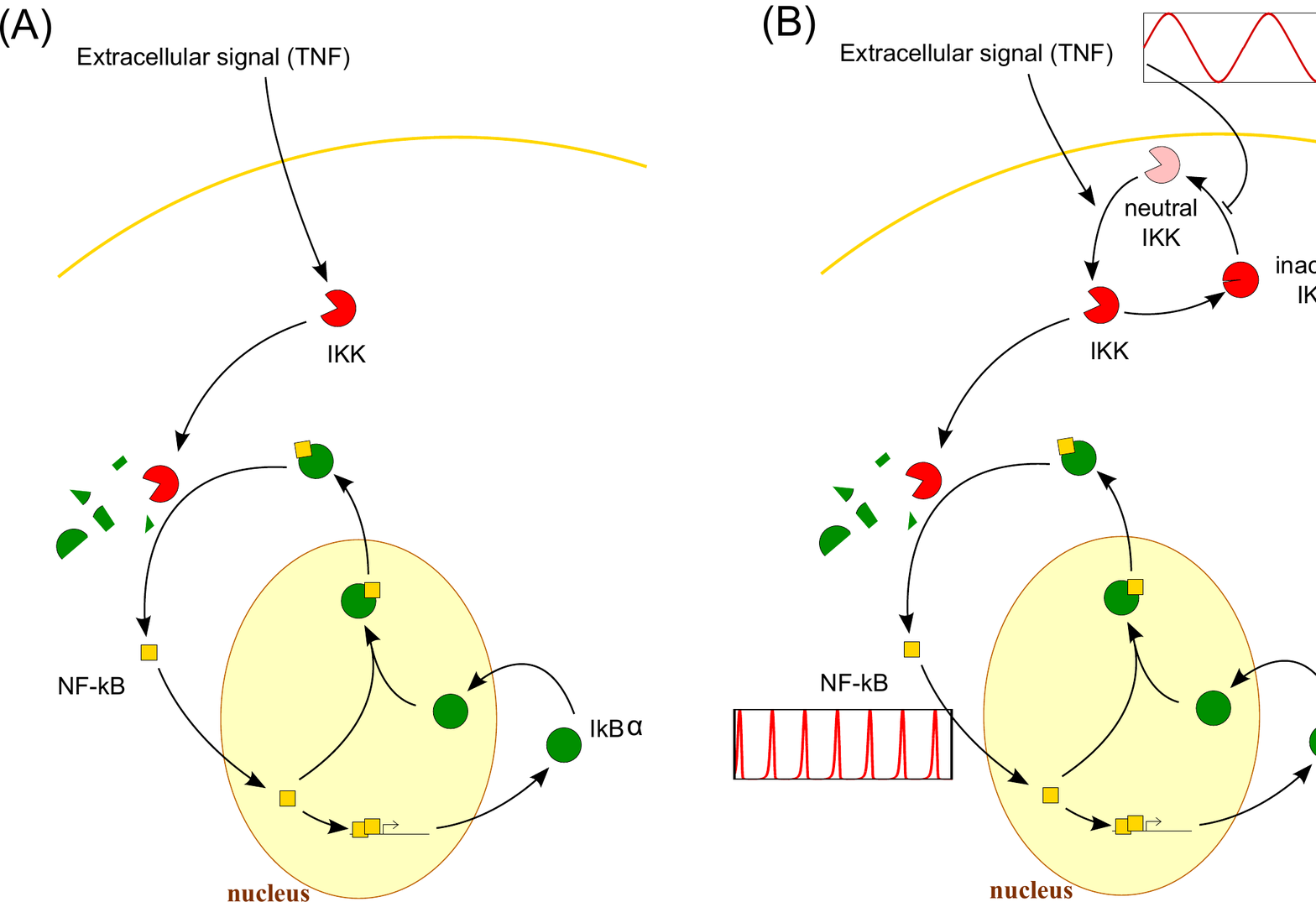}
\caption{(A) Schematic diagram of processes that form the core feedback loop
controlling NF-kB oscillations in the model of \cite{Krishna}. Nuclear NF-kB
activates transcription of IkB which sequesters NF-kB in the cytoplasm. IkB kinase (IKK),
when activated by external signals like the tumor necrosis factor alpha (TNF), causes
eventual targeted degradation of IkB when it is bound to NFkB. The released NF-kB
is then transported into the nucleus, closing the feedback loop. By assuming that
all complexes are in equilibrium, these processes can be represented by three
differential equations governing the dynamics of nuclear NFkB, IkB mRNA and cytoplasmic
IkB (see text). The external TNF signal's effect is simply modelled by appropriate choice of the
degradation rate of IkB. This is sufficient when examining only steady levels of TNF, as is
the case in \cite{Krishna}. (B) However, when examining the effect of temporally varying
TNF signals, this model has to be extended as shown to include the interaction between
TNF and various forms of IKK. We follow the model of \cite{Ashall} which consists of three
forms of IKK which inter-convert cyclically. TNF enhances the conversion of neutral to active
IKK, while also inhibiting the conversion of inactive to neutral IKK. Only active IKK is
able to cause degradation of IkB. These processes
can be represented by two additional differential equations (see text).}
\end{figure}

Here, we investigate the possibility of controlling the frequency
of ultradian cellular oscillators by synchronizing them to external
oscillations. Two important ultradian oscillators in mammalian cells
are triggered by external stresses. After
DNA-damage, the tumor suppressor protein p53 has been observed to oscillate with a
period of 4--5 hours \cite{Geva-Zatorsky06,Lahav08}. 
Secondly, inflammatory stresses
have been found to lead to oscillatory behavior in the transcription
factor NF-kB \cite{Nelson04}. Bulk and single cell
measurements after treatment with tumor necrosis factor (TNF)
show distinct and sharp oscillations with a time period of 2--3 hours \cite{Hoffmann02,Nelson04}. 

\begin{figure}[h!]
\includegraphics[width=0.8\textwidth]{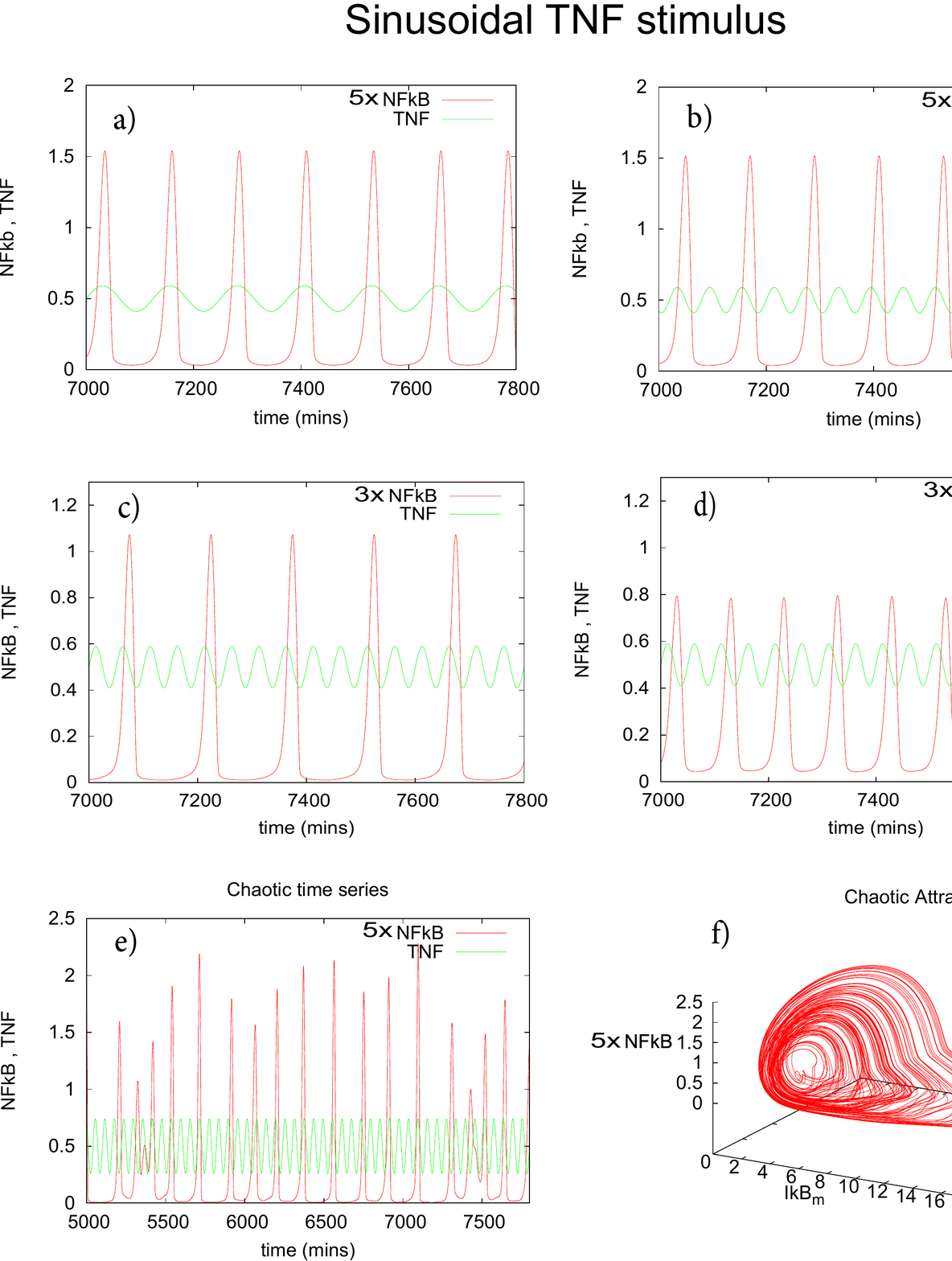}
\caption{(A) Simulation with applied TNF amplitude $A=0.09$ and frequency 
$\nu=1/2.08~hr^{-1}$ (which is close to the
reference frequency, $\nu_0=1/1.8~hr^{-1}$). Resultant NF-kB oscillations have the same frequency
as the applied TNF signal, giving the 1/1 ``tongue" in Fig. \ref{fig3}. 
(B) Simulation with $A=0.09$, $\nu=1/1.0~hr^{-1}$ 
(which is close to
twice the reference frequency). Resultant NF-kB oscillations have twice
the frequency of the applied TNF signal, giving the 1/2 ``tongue" in Fig. \ref{fig3}. 
(C) Simulation with $A=0.09$, $\nu=1/0.83~hr^{-1}$ with initial conditions $N_n=1~\mu M,~I_m=0.5 \mu M$ and all others zero $\mu M$. 
The NF-kB frequency synchronizes
to be 1/3 of the applied frequency.
(D) Simulation with $A=0.09$, $\nu=1/0.83~hr^{-1}$ identical to (C) except with different initial conditions where $N_n=0.1~\mu M,~I_m=0.5 \mu M$ and all others zero $\mu M$.
With these initial conditions, the NF-kB frequency synchronizes
to 1/2 of the applied frequency. Thus, for this value of $A$ and $\nu$, we observe multiple stable
synchronized states.
(E) Simulation with $A=0.24$, $\nu=1/1.0~hr^{-1}$.
At such large amplitudes chaotic oscillations are observed. (F) A plot
of the trajectory of oscillations in the configuration space. The shape of the trajectory
is quite typical of such chaotic oscillations and is known as a ``strange attractor" \cite{Strogatz}.}
\label{fig2}
\end{figure}

In these two cases, oscillations are caused by a feedback loop which incorporates the formation of a complex
between the transcription factor and an inhibitor (Mdm2 in the case of p53, and
IkB$\alpha$ in the case of NF-kB). The complex
formation induces an effective time delay through non-linear degradation 
which suffices to generate oscillations in 
the transcription factor and its inhibitor, out of phase with
each other \cite{Mengel}. Those two feedback loops have been modeled both by
applying explicit time delays \cite{Tiana} and by modelling the complex 
formation \cite{Mengel}.
An elaborate model with 26 variables (mRNAs, proteins, complexes, etc.) for the NF-kB system
was first formulated in Ref. \cite{Hoffmann02}.
Krishna et al. reduced this model to the core
feedback loop by assuming that complexes were in equilibrium \cite{Krishna}.
Fig. 1A shows a schematic representation of the resulting model that
consists of three coupled non-linear
differential equations: 
\begin{equation}\frac{dN_n}{dt} = k_{Nin}(N_{tot}-N_n)\frac{K_I}{K_I+I}-k_{Iin}I\frac{N_n}{K_N+N_n}\end{equation}
\begin{equation}\frac{dI_m}{dt} = k_tN_n^2-\gamma_mI_m\end{equation}
\begin{equation}\frac{dI}{dt} = k_{tl}I_m-\alpha [IKK]_a(N_{tot}-N_n)\frac{I}{K_I+I}\end{equation}

The triggering stimulus, e.g. TNF, acts by changing the level of active IkB
kinase, $[IKK]_a$, which phosphorylates IkB, resulting eventually in its degradation. This degradation rate
is one of the parameters of the model and Ref. \cite{Krishna} used different constant
values of this parameter to represent different steady levels of the TNF stimulus.
Default parameter values are given in Table 1. With these values and choosing
$[IKK]_a=0.5~\mu M$ as in \cite{Krishna}, one obtains sustained oscillations with
a frequency $\nu=1/0.9~hr^{-1}$.

Here we wish to examine the effect of an oscillatory TNF stimulus on the
system. 
The simplest possibility is to assume the degradation rate of IkB would oscillate identically
to the TNF stimulus,
and Ref. \cite{Fonslet} showed that this could result in chaotic NF-kB oscillations.
However, unlike steady levels of TNF,
representing an oscillatory TNF signal by a similarly
shaped oscillatory behavior of the IkB degradation rate is unjustifiable --
non-linear interaction between TNF and IKK
could well cause complex changes in the shape of the external signal
as it is transduced. Therefore, we extended the model of Krishna et al \cite{Krishna}
to include the circuit that transduces the TNF signal to the IKK concentration,
as shown schematically in Fig. 1B.
Ashall et al \cite{Ashall} have modelled this circuit in detail, and we add the
two relevant differential equations from their model to the Krishna et al model:

\begin{equation}\frac{d[IKK]_a}{dt} = k_a[TNF]([IKK]_{tot}-[IKK]_a-[IKK]_i)-k_i[IKK]_a\end{equation}
\begin{equation}\frac{d[IKK]_i}{dt} = k_i[IKK]_a-k_p[IKK]_i\frac{k_{A20}}{k_{A20}+[A20][TNF]}\end{equation}

This model assumes that there is a constant pool
of IKK which is interconverted between different
states -- active, inactive and neutral. TNF increases the rate at which inactive IKK is made
active, and it is only the active IKK which
phosphorylates IkB and thereby affects the degradation rate of IkB$\alpha$.
Ref. \cite{Ashall} constructed these equations so that the
TNF signal can be represented by a dimensionless number between 0 (off) and 1 (on).
We have used the parameters values from \cite{Ashall} (see Table 1) except for
$N_{tot}$ and $[IKK]_{tot}$. Ref. \cite{Ashall} takes both $\sim 0.1\mu M$, whereas
\cite{Krishna} uses $N_{tot}=1\mu M$. Here, we chose to keep $N_{tot}=1\mu M$
and varied $[IKK]_{tot}$ around 1$\mu M$, with $[TNF]$ fixed at 0.5, to find a value
that gave sustained spiky oscillations with a frequency in the range 0.3--1 $hr^{-1}$.
The model in \cite{Ashall} also includes another
slow feedback via the molecule A20 as seen in the equations.
For simplicity we ignore this feedback by keeping its concentration, $[A20]$, constant
as this feedback loop mainly fine-tunes the shape of NF-kB response and
``there is a range of constitutive A20 expression values that can functionally replace A20 negative feedback" \cite{Werner}.
As default values, we finally chose the combination of $[IKK]_{tot}=2~\mu M$ and $[A20]=0.0026~\mu M$.
This results in sustained spiky oscillations of frequency $\nu_0\approx 1/1.8~hr^{-1}$
when $[TNF]$ is fixed at 0.5. 
Below, we show that with these parameter values oscillatory TNF stimuli in the model defined by equations (1)--(5)
can produce both very organized responses -- a multitude
of synchronized states -- as well as chaotic behavior.
We have checked that our results are not qualitatively changed by perturbation
of parameters around these default values, as long as the parameters result in
sustained spiky oscillations when TNF is kept fixed at 0.5 (data not shown).

We examine both sinusoidal as well as square wave oscillation of TNF, and
in each case vary the amplitude and frequency of the applied stimulus, while
keeping the rest of the parameters of the model fixed.
When TNF is sinusoidally varied with
different frequencies around the average value 0.5, we observe that
the NF-kB oscillation synchronizes in interesting ways to the applied TNF frequency.
For example, when the applied frequency is close to the $\nu_0$, then
the NF-kB oscillation synchronizes to have the same frequency as the external
stimulus, as well as a constant phase difference between the maxima of TNF and NF-kB (see Fig 2a).
Similarly, when the applied frequency is in a band close to $2\nu_0$, the NF-kB
oscillation synchronizes to a period exactly twice the applied frequency (see Fig. 2b).
A fundamental result of our investigation is that {\it the NF-kB oscillations
will stay completely synchronized even if the frequency of TNF oscillations is slightly diminished
or slightly increased!}. That is, the external TNF signal is able
to `pull' the frequency of the NF-kB oscillation towards a rational ratio with respect to
the applied frequency. This is known as phase (or mode) locking \cite{JBB}. 
As the amplitude of the applied oscillation increases, these bands
of synchronization expand and the resulting shapes are called
Arnold tongues \cite{JBB}. Fig. 3a shows the Arnold tongues for the
case where the TNF signal is sinusoidal, 
$[TNF](t)=0.5+Asin(2\pi\nu t)$,
with varying frequency, $\nu$, and amplitude, $A$.
Fig. 3b shows a similar diagram for the case when TNF is a square wave:
$[TNF](t)=0.5+A[sign(sin(2\pi\nu t)+1]/2$. This protocol may be easier to
realize experimentally than a sinusoidal one, and in fact shows broader
Arnold tongues.

In principle, there is an Arnold tongue wherein NF-kB shows $p$ peaks for every $q$ peaks of TNF,
for every rational number $p/q$ (where $p$ and $q$ are natural numbers).
The width of each tongue starts off infinitely
small when $A=0$, and expands smoothly as $A$ increases. Evidently, this cannot happen
without tongues overlapping, and indeed such overlaps occur as soon as $A>0$.
In general, in overlapping regions one expects to observe multistability, i.e.
multiple synchronized states with different $p/q$ values will coexist, with different
states being realized when different initial conditions are used.
However, for small $A$, the states corresponding to small $p$ and $q$ numbers
generally dominate the observed behavior. As $A$ is increased, these dominant
states, such as 1/1 and 2/1, also start overlapping and then one can actually
observe multistability (see Figs. 2c,d). As $A$ is increased further, and there are more
and more overlaps, one can also encounter chaotic behavior as shown in Figs. 2e,f.
The same behavior occurs for square wave oscillations of TNF (data not shown).

This complex behavior of the existence, growth and overlapping of
Arnold tongues is observed in several very simple sets of
nonlinear differential equations, such as circle maps and other return maps
(we refer the reader to \cite{JBB, Kurths, He} for details).
It has also been observed in a number of physical systems ranging from
turbulent fluids, where synchronized states with rational
numbers up to 83/79 haven been measured \cite{Stavans}, quantum mechanical devices
like Josephson junctions and semi-conductors \cite{alstrom,He,Gwinn,Lindsey},
crystals \cite{Martin}, and sliding charge-density waves \cite{Gruner}.
Synchronization is known to occur in living systems,
such as fireflies, and circadian clocks entrain to the day-night cycle \cite{Asher}. 
However, to our knowledge such Arnold tongues have not been
observed {\em in vivo} at a subcellular level. 

Our work suggests that this kind of
intricate synchronization could be observed in the NF-kB
system, and also the p53-Mdm2 system as it has a very similar core feedback
loop.
More specifically, we predict that:
(a) oscillations in NF-kB can be synchronized to TNF oscillations, (b) the bigger the amplitude,
the stronger the synchronization (when amplitudes are relatively small), (c) the oscillations can in principle be
synchronized to all rational ratios with respect to the applied frequency,
but states with smaller $p$ and $q$ values will dominate, (d) if oscillations
can be sustained for around a day, the states 1/2, 1/1 and 2/1 should be
observable in practice, (e) when the amplitude of TNF oscillations is increased
further, chaotic behavior will appear. 
Similar predictions could be made for the p53-Mdm2 system, 
controlled by external stimuli such as irradiation or exposure to DNA-damaging
chemicals. If this basic synchronization technique works, we hope it could be
developed into a tool to control inflammatory and apoptotic pathways
and ultimately to regulate DNA repair, immune response and 
eventually cell fate.  

\begin{figure}[h!]
(A,B)\\
\includegraphics[width=.6\textwidth]{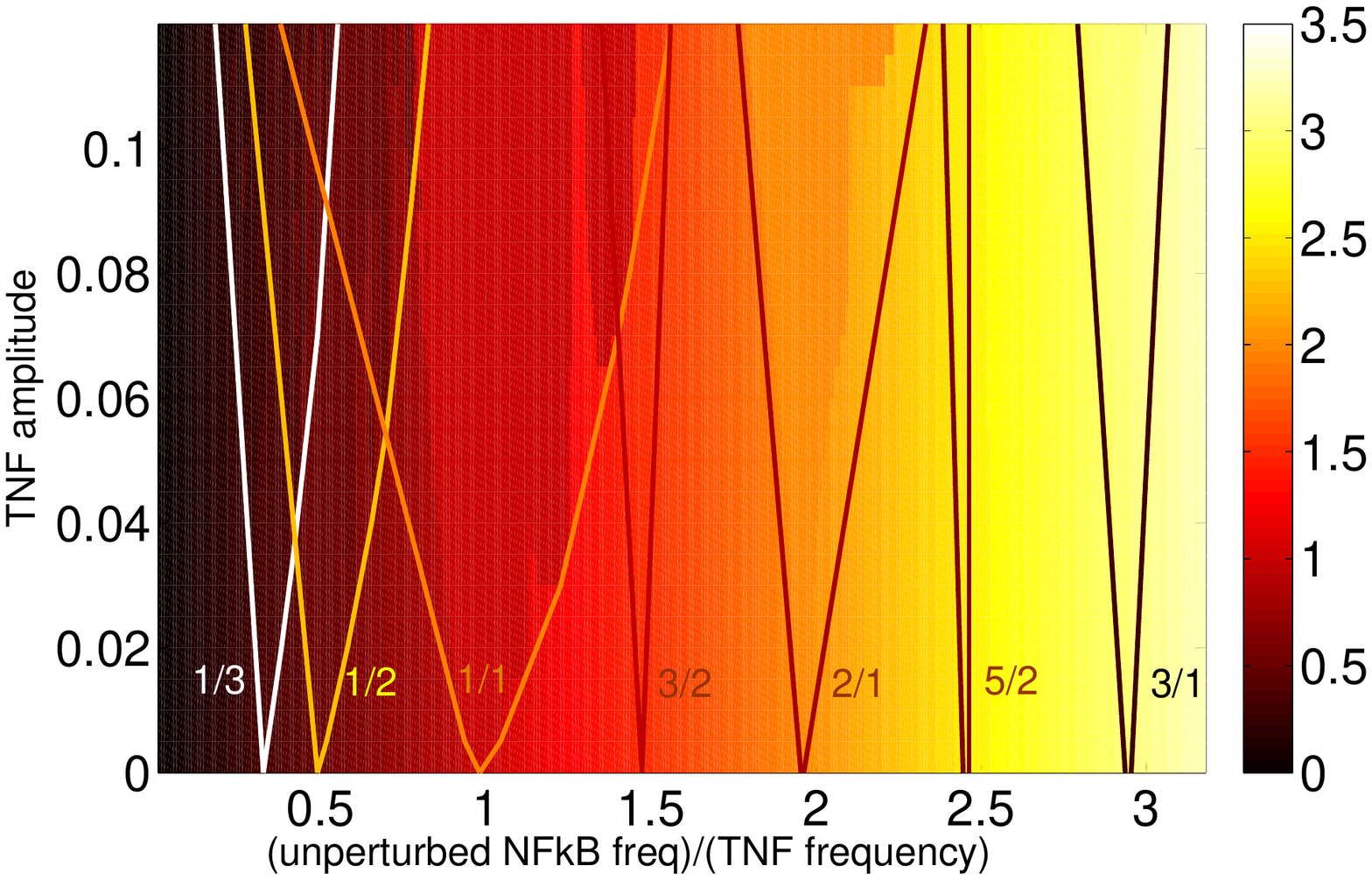}
\includegraphics[width=.6\textwidth]{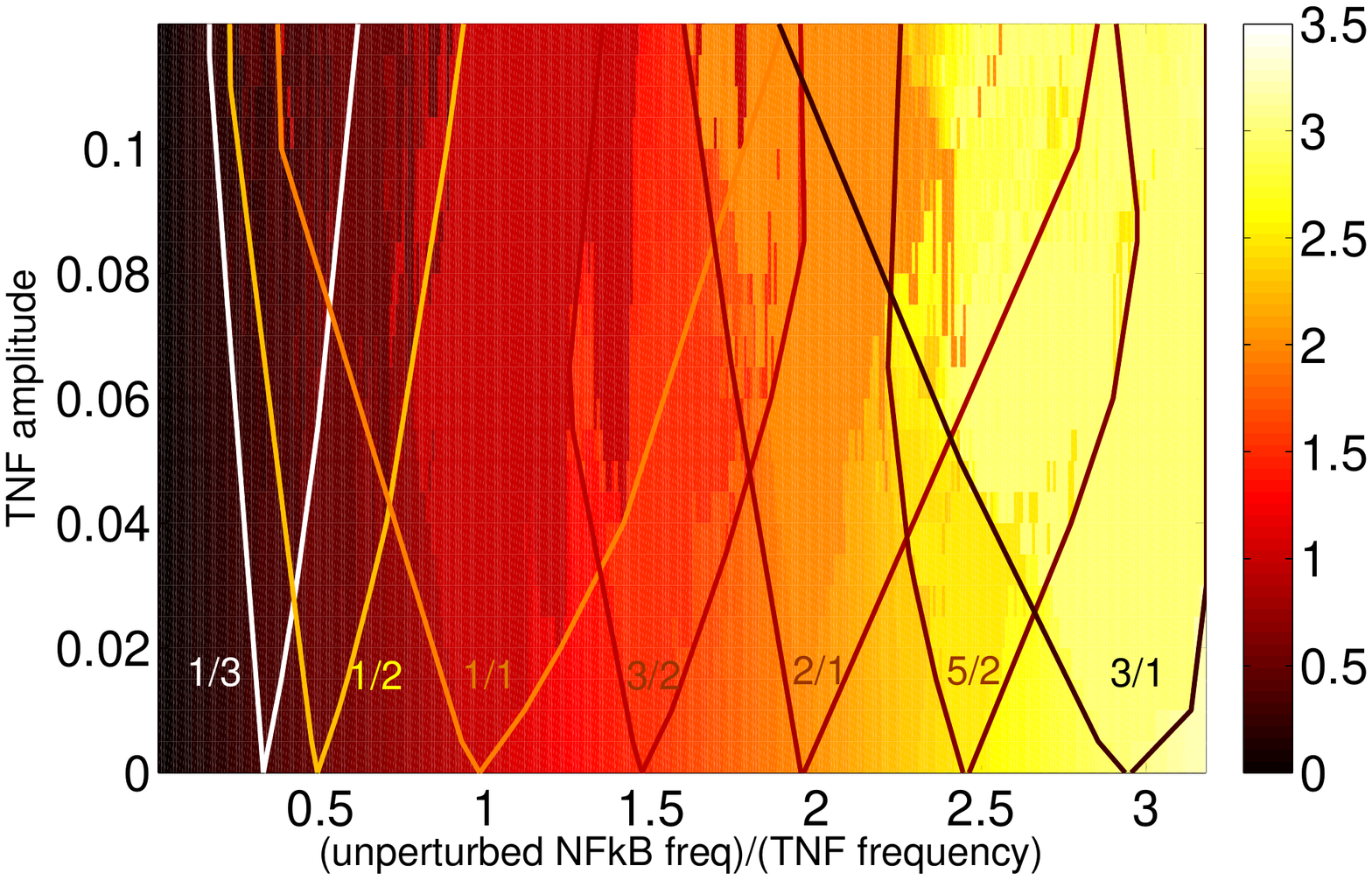}
\caption{(A) Colors show the ratio of observed NF-kB frequency to the applied TNF
frequency, as a function of TNF frequency and amplitude. The applied TNF signal
is sinusoidal: $[TNF]=0.5+Asin(2\pi\nu t)$. The Arnold tongues corresponding to the
states 1/3,1/2,1/1,3/2,2/1,5/2,3/1 are shown from left to right. For a tongue
$p/q$, the boundary shown is the convex hull of points in the parameter space where the ratio of applied to
observed frequency was within 0.5\% of $p/q$. (B) Similar Arnold tongue diagram
for the case where a square wave TNF signal is applied: $[TNF]=0.5+A[sign(sin(2\pi\nu t))+1]/2$.}
\label{fig3}
\end{figure}

%

{\bf Acknowledgement}. This work was supported by the Danish National Science Foundation through
the ``Center for Models of Life". We are grateful to Markus Covert for 
discussions on NF-kB oscillations and possible mode-locking in cells.

\begin{table}
\begin{tabular}{|l|l|}
Parameter & Default value \\
\hline
$k_{Nin}$ & 5.4 min$^{-1}$\\
$k_{Iin}$ & 0.018 min$^{-1}$\\
$k_t$ & 1.03 ($\mu$M)$^{-1}$.min$^{-1}$\\
$k_{tl}$ & 0.24 min$^{-1}$\\
$K_I$ & 0.035 $\mu$M\\
$K_N$ & 0.029 $\mu$M\\
$\gamma_m$ & 0.017 min$^{-1}$\\
$\alpha$ & 1.05 ($\mu$M)$^{-1}$.min$^{-1}$\\
$N_{tot}$ & 1. $\mu$M\\
\hline
$k_a$ & 0.24 min$^{-1}$\\
$k_i$ & 0.18 min$^{-1}$\\
$k_p$ & 0.036 min$^{-1}$\\
$k_{A20}$ & 0.0018 $\mu$M\\
\hline
$[IKK]_{tot}$ & 2.0 $\mu$M\\
$[A20]$ & 0.0026 $\mu$M\\
\end{tabular}
\caption{Default values of parameters in the model. The first 9 are from Ref. \cite{Krishna} and 
the next 4 from Ref. \cite{Ashall}. $[IKK]_{tot}$ and $[A20]$ were chosen
in order to obtain sustained spiky oscillations with frequency in the range 0.3--1 $hr^{-1}$ when 
$[TNF]$ is kept fixed at 0.5 (the actual frequency obtained with these values is $\nu_0=1/1.8~hr^{-1}$.)}
\end{table}

\end{document}